\documentclass[10pt,draft]{article}
\usepackage{amsfonts,amsmath,amssymb,cite}

\usepackage{pdflscape}
\begin{document}
\title{Para- and diamagnetic contributions to magnetic shielding constants of relativistic hydrogenlike atoms in some low-lying discrete energy eigenstates}
\author{Patrycja Stefa{\'n}ska\footnote{Corresponding author. 
Email: patrycja.stefanska@pg.edu.pl} \\*[2ex]
Institute of Physics and Computer Science, \\
Faculty of Applied Physics and Mathematics, \\
Gda{\'n}sk University of Technology,
Narutowicza 11/12, 80--233 Gda{\'n}sk, Poland}
\date{}
\maketitle
\begin{abstract}
We present tabulated data for numerical calculations of relative para- and diamagentic contributions to the magnetic shielding constant ($\sigma$) of the Dirac one-electron atoms with a pointlike, spinless and motionless nuclei of charge $Ze$. Exploiting the analytical formulas for the diamagnetic ($\sigma_{d}$) and paramagnetic ($\sigma_{p}$) terms of $\sigma$, valid for an arbitrary discrete energy state, recently derived by us with the aid of the Gordon decomposition technique, we have found the numerical values of $\sigma_{d}/\sigma$ and $\sigma_{p}/\sigma$ for the ground state and for the first two sets of excited states (i.e.: $2s_{1/2}$, $2p_{1/2}$, $2p_{3/2}$, $3s_{1/2}$, $3p_{1/2}$, $3p_{3/2}$, $3d_{3/2}$, and $3d_{5/2}$) of the relativistic hydrogenic ions with the nuclear charge number from the range $1 \leqslant Z \leqslant 137$.   The comparisons of our results with those reported by other authors for some atomic states are also presented.  We also compile here the numerical values of the total magnetic shielding constants for the ground state $1s_{1/2}$ and for each state belonging to the first set of excited states of selected hydrogenlike ions, obtained with the use of three different values of the fine-structure constant, i.e.:  $\alpha^{-1}=137.035 \: 999 \: 139$ (from CODATA 2014), $\alpha^{-1}=137.035 \: 999 \: 084$ (from CODATA 2018) and $\alpha^{-1}=137.035 \: 999 \: 177$ (from CODATA 2024).
\end{abstract}
%
%
\section{Introduction}
\label{I}
\setcounter{equation}{0}
A few years ago, we have launched a research program for the systematic calculations of various electromagnetic susceptibilities of Dirac one-electron atoms. The analitycal part of our work consisted in deriving closed-form expressions for some atomic parameters characterizing the response of the relativistic hydrogenlike atom to the external perturbations. These calculations were made using the Sturmian expansion of the generalized Dirac--Coulomb Green function (proposed in Ref.\ \cite{Szmy97}) combined with the theory of hypergeometric functions and gave interesting results for the atomic ground state \cite{Szmy11, Stef12, Szmy12, Szmy14} and for an arbitrary excited states of the atom \cite{Stef15, Stef16a, Stef16b}. Because most of the derived expressions, among others, for the magnetizability and for the cross-susceptibilitiy are of fairly complex form, we have systematically programmed them in order to obtain the numerical values for theese physical quantities; the numerical parts of our results were published in Refs.\ \cite{Stef16, Stef17, Stef18}.

Following the suggestions from one of Referees of our paper \cite{Stef16b}, we started to study the diamagnetic and paramagnetic contributions to some magnetic susceptibilities of the relativistic hydrogenlike atoms being in an arbitrary excited states.  The first subject of our new research was magnetizability \cite{Stef24, Stef20a}, as one of the most commonly used parameter in atomic physics \cite{Vleck32}. The second most important quantity in physics and chemistry is the magnetic shielding constant ($\sigma$) of the atomic nucleus, also used in medicine, e.g. to the analyzing of the signal in the NMR (Nuclear Magnetic Resonance)  spectra. To derive the analytical expressions for dia- and paramagnetic contributions to $\sigma$, as in the case of the magnetizability, we have exploited the Gordon decomposition approach \cite{Gord28}. After quite complex calculations, which used, among others, the Sturmian expansion of the Dirac--Coulomb Green function \cite{Szmy97} and the theory of special functions, we have found the atomic magnetic shielding constant ($\sigma$) as a sum
\begin{equation}
\sigma=\sigma_{d}+\sigma_{p},
\label{eq:1}
\end{equation}
in which the diamagnetic ($\sigma_{d}$) and paramagnetic ($\sigma_{p}$) contributions to $\sigma$ are obtained as the following functions of the three quantum numbers $\{n, \kappa, \mu \}$, describing the unperturbed atomic state ($n$ is the radial quantum number, \mbox{$\kappa=\pm1, \pm2, \ldots$} is the Dirac quantum number, while $\mu$ stands for the magnetic quantum number): 
\begin{equation}
\sigma_d
=\frac{\alpha^2 Z}{4 N_{n \kappa}^2}  \left(1+\frac{4\mu^2}{4\kappa^2-1}\right) 
\label{eq:2}
\end{equation}
and
\begin{eqnarray}
\sigma_p=\frac{\alpha^2 Z}{N_{n\kappa}^2(4\kappa^2-1)} 
\Bigg\{ 
\frac{1-\eta_{n\kappa}^{(+)}-\eta_{n\kappa}^{(-)}}{4} - \frac{16\kappa^3\mu^2}{4\kappa^2-1} \frac{(\alpha Z)^4}{\gamma_{\kappa}(4\gamma_{\kappa}^2-1)N_{n\kappa}^3}\left(\eta_{n\kappa}^{(+)}+\eta_{n\kappa}^{(-)} \right)
\nonumber \\ 
 +\frac{\mu^2}{(4\kappa^2-1)^2}
\left[
(2\kappa+1)^2\eta_{n\kappa}^{(+)}+(2\kappa-1)^2\eta_{n\kappa}^{(-)}-12\kappa^2-1
\right]
\Bigg\},
\label{eg:3}
\end{eqnarray}
where we have defined
\begin{equation}
\eta_{n\kappa}^{(\pm)}=\frac{(2\kappa \pm 1)N_{n\kappa}}{n+\gamma_{\kappa}\pm N_{n\kappa}}.
\label{eq:4}
\end{equation}
Here, $\alpha$ denotes the Sommerfeld's fine structure constant, $a_0$ is the Bohr radius, whereas
\begin{equation}
N_{n\kappa}=\sqrt{n^2+2n\gamma_{\kappa}+\kappa^2},
\label{eq:5}
\end{equation}
and
\begin{equation}
\gamma_{\kappa}=\sqrt{\kappa^2-(\alpha Z)^2}.
\label{eq:6}
\end{equation}
With the use of Eqs.\ (\ref{eq:2})--(\ref{eq:6}), we have checked that the sum $\sigma_{d}+\sigma_{p}$ agrees with the closed-form expression for the total magnetic shielding constant of the relativistic hydrogenic atoms, derived by us some time ago \cite{Stef16b}.  Moreover, after redefining several parametres and performing further some algebraic transformations, one can show that the above general results, written for some low-lying energy states, coincide with the corresponding formulas reported in Refs.\ \cite{Pype99, Pype99a}.

However, what researchers usually need the most are numerical values of the atomic and molecular parameters, which can be used in many different fields of science. Therefore, as well as due to the lack of such data in the scientific literature, we performed numerical calculations for the relative dia- and paramagnetic contributions to $\sigma$ for some selected atomic states.  In the next section we will discuss briefly our results.

%
%
\section{Discussion of results}
\label{II}
\setcounter{equation}{0}
All the numerical results presented in this work have been obtained with the help of the formulas from Eqs.\ (\ref{eq:1})--(\ref{eq:6}).  In Table \ref{tab:1s_1-2} we have included the values of the \emph{relative} dia- and paramagnetic contributions to the magnetic shielding constants of the relativistic hydrogenlike atoms being in the ground state $1s_{1/2}$. Tables \ref{tab:2s_1-2}--\ref{tab:3d_5-2_mu_5-2} contain the corresponding results for states belonging to the first two sets of excited states, i.e.: $2s_{1/2}$, $2p_{1/2}$, $2p_{3/2}$, $3s_{1/2}$, $3p_{1/2}$, $3p_{3/2}$, $3d_{3/2}$ and $3d_{5/2}$, having regarded all possible values of the magnetic quantum number $\mu$. The reader should observe that for the states with $|\kappa|=1$, there is a limitation for the nuclear charge number, i.e.: $Z<\alpha^{-1}\frac{\sqrt{3}}{2} \simeq 118.67$ (see Tables \ref{tab:1s_1-2}--\ref{tab:2p_1-2}, \ref{tab:3s_1-2}, and \ref{tab:3p_1-2}); a detailed explanation of this restriction can be found in Ref.\ \cite{Stef16b}, just follow Eq.\ (3.17). 

The value of the inverse of the fine-structure constant used during creation of Tables \ref{tab:1s_1-2}--\ref{tab:3d_5-2_mu_5-2} was $\alpha^{-1}=137.035 \: 999 \: 177$,  and was taken from the newest  CODATA report on Recommended Values of the Fundamental Physical Constants \cite{Nist24}.  However, in order to make the best comparison of the present results with the numerical results reported earlier by other authors for some atomic states \cite{Pype99a}, we have performed two additional tables (Tables \ref{tab:comparison2}--\ref{tab:comparison3}) with the values of $\sigma_d$ and $\sigma_p$ obtained using $\alpha^{-1}=137.035 \: 98 \:95$ (from CODATA 1986) \cite{Cohe88}. To show how the change in the value of the fine-structure constant affects the value of the \emph{total} magnetic shielding constants, additionally, we performed calculations for $\sigma$ with the current value of $\alpha^{-1}$ and also with its two previous values: $\alpha^{-1}=137.035 \: 999 \: 084$ (from CODATA 2018) \cite{Nist18} and  $\alpha^{-1}=137.035 \: 999 \: 139$ (from CODATA 2014) \cite{Mohr14}; the appropriate juxtaposition of these results for the ground state and for the first set of excited states for selected hydrogenlike ions is presented in Table \ref{tab:comparison}.

It was already stated in the previous paragraph, that some representative results for (not relative!) dia- and paramagnetic contributions to the magnetic shielding constants of the relativistic one-electron atoms can be found in the paper of Pyper and Zhang \cite{Pype99a}. Since they introduced a paramagnetic term $\sigma_p$ in the form of the sum of two components, we had to sum them up to be able to compare the result with our corresponding values for $\sigma_p$. The agreement between these results, as well as for the data for $\sigma_d$ turns to be almost perfect, but it should be noted that our results are of much greater accuracy. Thus, in some cases, e.g. for $Z=11$, the differences may by treated as the results of numerical rounding.

Tables \ref{tab:1s_1-2}--\ref{tab:3d_5-2_mu_5-2} show that for states with the angular-plus-parity symmetry quantum number $\kappa<0$ and simultaneously with the maximal value of $\mu$ (i.e. when the magnetic quantum number $\mu$ is equal to the total angular momentum quantum number $j$), we have $\sigma_d>\sigma_p$. But for the $s$-states such relations occurs not for the whole range of atomic number: it is most evident for small $Z$'s and gradually decreases to a certain $Z_c$ value, for which both contributions are almost equal, and then the paramagnetic component slowly begins to dominate over the diamagnetic one. For the ground state we have found $Z_c=67$, whereas for the highest excited states the value of $Z_c$ increases: for state $2s_{1/2}$ we have $Z_c=79$, and for state $3s_{1/2}$ there is  $Z_c=86$.

The inverse relation, i.e. when $\sigma_d<\sigma_p$, occurs for those states, for which $\kappa<0$ and $\mu<j$ simultaneously, and also for states with $\kappa>0$. Moreover, for the first of the listed classes of atomic states, the relative diamagnetic contribution to $\sigma$ is negative. For the latter class of states, however, we have a interesting behavior of the paramagnetic contribution. Namely, for states with $\kappa=+1$ (the $p_{1/2}$-states), $\sigma_p/\sigma$ slowly decreases up to a certain $Z_c$ value, after which it starts to increase slightly. We have found $Z_c=99$ for $2p_{1/2}$ and  $Z_c=105$ for $3p_{1/2}$.

\section{Acknowledgment}
I am grateful to the anonymous Referee of my paper \cite{Stef16b} for his/her suggestion to make and publish present work.

\newpage

\section{Table explanation}
\label{V}
\setcounter{equation}{0}
In all the tables we have used the following notation:

\begin{tabular}{ll}
$\sigma$ 				& the magnetic shielding constant \\
$\sigma_{d}$ 		& the diamagnetic part of the magnetic shielding constant \\
$\sigma_{p}$ 		& the paramagnetic part of the magnetic shielding constant \\
$Z$ 					& the nuclear charge number \\
$\alpha$ 			& the fine-structure constant \\
$a_0$					& the Bohr radius \\
$a(b)$				& this means $a \times 10^b$ \\
$\mu$         & the magnetic quantum number; $\mu=-j, \ldots, j$ \\
\end{tabular}
\\
\noindent
The atomic states are referred to as $\mathcal{N}{x}_{j}$, due to the following guidelines:

\begin{tabular}{ll}
$\mathcal{N}$  & the principal quantum number \\
$\mathcal{N}=n+|\kappa|$ & Note that for the radial quantum number $n=0$, $\kappa$ is the negative integer! \\
$\kappa$ 			& the Dirac quantum number; $\kappa=(l-j)(2j+1)$\\
$j$						& the total angular momentum quantum number; $j=|\kappa|-\frac{1}{2}=l \pm \frac{1}{2}$ \\
$x \equiv s$	& states with the orbital angular momentum quantum number $l=0$ \\
$x \equiv p$	& states with the orbital angular momentum quantum number  $l=1$ \\
$x \equiv d$	& states with the orbital angular momentum quantum number  $l=2$ \\
\end{tabular}

\section{Data Tables}
\label{V}
\setcounter{equation}{0}

\begin{table}
\footnotesize
\caption{\small{Relative dia- and paramagnetic contributions to magnetic shielding constants for the ground state  ($1s_{1/2}$) of hydrogenlike atoms, obtained with $\alpha^{-1}=137.035 \: 999 \: 177$ (CODATA 2024).}}
\label{tab:1s_1-2}
\begin{tabular}{rccrcc}
\hline
$Z$ & $\sigma_{d}/ \sigma$ & $\sigma_{p}/ \sigma$ & Z  & $\sigma_{d}/ \sigma$ &  $\sigma_{p}/ \sigma$ \\
\hline
1	&	9.998565263884($-1$)	&	1.434736115899($-4$)	&	60	&	5.819500035373($-1$)	&	4.180499964627($-1$)	\\
2	&	9.994262159823($-1$)	&	5.737840177483($-4$)	&	61	&	5.707128831581($-1$)	&	4.292871168419($-1$)	\\
3	&	9.987093998715($-1$)	&	1.290600128541($-3$)	&	62	&	5.594438550715($-1$)	&	4.405561449285($-1$)	\\
4	&	9.977066292203($-1$)	&	2.293370779737($-3$)	&	63	&	5.481478102959($-1$)	&	4.518521897041($-1$)	\\
5	&	9.964186742910($-1$)	&	3.581325708959($-3$)	&	64	&	5.368295730837($-1$)	&	4.631704269163($-1$)	\\
6	&	9.948465230824($-1$)	&	5.153476917618($-3$)	&	65	&	5.254938997023($-1$)	&	4.745061002977($-1$)	\\
7	&	9.929913795860($-1$)	&	7.008620414002($-3$)	&	66	&	5.141454774412($-1$)	&	4.858545225588($-1$)	\\
\hline
\end{tabular}
\end{table}

\begin{table}
\footnotesize
\caption{\small{Relative dia- and paramagnetic contributions to magnetic shielding constants  of hydrogenlike atoms in the excited state $2s_{1/2}$, obtained with $\alpha^{-1}=137.035 \: 999 \: 177$ (CODATA 2024).}}
\label{tab:2s_1-2}
\begin{tabular}{rccrcc}
\hline
$Z$ & $\sigma_{d}/ \sigma$ & $\sigma_{p}/ \sigma$ & Z  & $\sigma_{d}/ \sigma$ &  $\sigma_{p}/ \sigma$ \\
\hline
1	&	9.999286263903($-1$)	&	7.137360968616($-5$)	&	60	&	7.211318366960($-1$)	&	2.788681633040($-1$)	\\
2	&	9.997144815321($-1$)	&	2.855184679358($-4$)	&	61	&	7.111840347461($-1$)	&	2.888159652539($-1$)	\\
3	&	9.993574933602($-1$)	&	6.425066397897($-4$)	&	62	&	7.010494433697($-1$)	&	2.989505566303($-1$)	\\
4	&	9.988575418416($-1$)	&	1.142458158414($-3$)	&	63	&	6.907288508082($-1$)	&	3.092711491918($-1$)	\\
5	&	9.982144590882($-1$)	&	1.785540911808($-3$)	&	64	&	6.802232494757($-1$)	&	3.197767505243($-1$)	\\
6	&	9.974280295167($-1$)	&	2.571970483309($-3$)	&	65	&	6.695338488354($-1$)	&	3.304661511646($-1$)	\\
7	&	9.964979900550($-1$)	&	3.502009945010($-3$)	&	66	&	6.586620886220($-1$)	&	3.413379113780($-1$)	\\
\hline
\end{tabular}
\end{table}

\begin{table}
\footnotesize
\caption{\small{Relative dia- and paramagnetic contributions to magnetic shielding constants  of hydrogenlike atoms in the excited state $2p_{1/2}$, obtained with $\alpha^{-1}=137.035 \: 999 \: 177$ (CODATA 2024).}}
\label{tab:2p_1-2}
\begin{tabular}{rccrcc}
\hline
$Z$ & $\sigma_{d}/ \sigma$ & $\sigma_{p}/ \sigma$ & Z  & $\sigma_{d}/ \sigma$ &  $\sigma_{p}/ \sigma$ \\
\hline
1	&	2.995358778603($-5$)	&	9.999700464122($-1$)	&	60	&	1.030316723278($-1$)	&	8.969683276722($-1$)	\\
2	&	1.198107614459($-4$)	&	9.998801892386($-1$)	&	61	&	1.062777235413($-1$)	&	8.937222764587($-1$)	\\
3	&	2.695607473383($-4$)	&	9.997304392527($-1$)	&	62	&	1.095578978006($-1$)	&	8.904421021994($-1$)	\\
4	&	4.791855742088($-4$)	&	9.995208144258($-1$)	&	63	&	1.128705079128($-1$)	&	8.871294920872($-1$)	\\
5	&	7.486600503049($-4$)	&	9.992513399497($-1$)	&	64	&	1.162137402715($-1$)	&	8.837862597285($-1$)	\\
6	&	1.077951731124($-3$)	&	9.989220482689($-1$)	&	65	&	1.195856432799($-1$)	&	8.804143567201($-1$)	\\
7	&	1.467020877578($-3$)	&	9.985329791224($-1$)	&	66	&	1.229841145697($-1$)	&	8.770158854303($-1$)	\\
\hline
\end{tabular}
\end{table}

\newpage
\thispagestyle{empty}

\begin{table}
\footnotesize
\caption{\small{Relative dia- and paramagnetic contributions to magnetic shielding constants  of hydrogenlike atoms in the excited state $2p_{3/2}$ ($\mu=\pm 1/2$), obtained with $\alpha^{-1}=137.035 \: 999 \: 177$ (CODATA 2024).}}
\label{tab:2p_3-2}
\begin{tabular}{rccrcc}
\hline
$Z$ & $\sigma_{d}/ \sigma$ & $\sigma_{p}/ \sigma$ & Z  & $\sigma_{d}/ \sigma$ &  $\sigma_{p}/ \sigma$ \\
\hline
1	&	-2.396384326641($-5$)	&	1.000023963843($0$)	&	70	&	-1.387009669676($-1$)	&	1.138700966968($0$)	\\
2	&	-9.586417947269($-5$)	&	1.000095864179($0$)	&	71	&	-1.434587462344($-1$)	&	1.143458746234($0$)	\\
3	&	-2.157274370686($-4$)	&	1.000215727437($0$)	&	72	&	-1.483376977964($-1$)	&	1.148337697796($0$)	\\
4	&	-3.835976942515($-4$)	&	1.000383597694($0$)	&	73	&	-1.533411333417($-1$)	&	1.153341133342($0$)	\\
5	&	-5.995367252031($-4$)	&	1.000599536725($0$)	&	74	&	-1.584725122876($-1$)	&	1.158472512288($0$)	\\
6	&	-8.636240649530($-4$)	&	1.000863624065($0$)	&	75	&	-1.637354497734($-1$)	&	1.163735449773($0$)	\\
7	&	-1.175957092999($-3$)	&	1.001175957093($0$)	&	76	&	-1.691337251927($-1$)	&	1.169133725193($0$)	\\
\hline
\end{tabular}
\end{table}

\clearpage

\thispagestyle{empty}

\begin{table}
\footnotesize
\caption{\small{Relative dia- and paramagnetic contributions to magnetic shielding constants  of hydrogenlike atoms in the excited state $2p_{3/2}$ ($\mu=\pm 3/2$), obtained with $\alpha^{-1}=137.035 \: 999 \: 177$ (CODATA 2024).}}
\label{tab:2p_3-2_mu_3-2}
\begin{tabular}{rccrcc}
\hline
$Z$ & $\sigma_{d}/ \sigma$ & $\sigma_{p}/ \sigma$ & Z  & $\sigma_{d}/ \sigma$ &  $\sigma_{p}/ \sigma$ \\
\hline
1	&	9.999714665214($-1$)	&	2.853347857858($-5$)	&	70	&	8.686047303850($-1$)	&	1.313952696150($-1$)	\\
2	&	9.998858705516($-1$)	&	1.141294483792($-4$)	&	71	&	8.650582948259($-1$)	&	1.349417051741($-1$)	\\
3	&	9.997432254867($-1$)	&	2.567745132728($-4$)	&	72	&	8.614741429925($-1$)	&	1.385258570075($-1$)	\\
4	&	9.995435536480($-1$)	&	4.564463519632($-4$)	&	73	&	8.578527420416($-1$)	&	1.421472579584($-1$)	\\
5	&	9.992868862736($-1$)	&	7.131137263626($-4$)	&	74	&	8.541945616049($-1$)	&	1.458054383951($-1$)	\\
6	&	9.989732635067($-1$)	&	1.026736493307($-3$)	&	75	&	8.505000736602($-1$)	&	1.494999263398($-1$)	\\
7	&	9.986027343804($-1$)	&	1.397265619607($-3$)	&	76	&	8.467697524049($-1$)	&	1.532302475951($-1$)	\\
\hline
\end{tabular}
\end{table}

\newpage

\begin{table}
\footnotesize
\caption{\small{Relative dia- and paramagnetic contributions to magnetic shielding constants  of hydrogenlike atoms in the excited state $3s_{1/2}$, obtained with $\alpha^{-1}=137.035 \: 999 \: 177$ (CODATA 2024).}}
\label{tab:3s_1-2}
\begin{tabular}{rccrcc}
\hline
$Z$ & $\sigma_{d}/ \sigma$ & $\sigma_{p}/ \sigma$ & Z  & $\sigma_{d}/ \sigma$ &  $\sigma_{p}/ \sigma$ \\
\hline
1	&	9.999524985525($-1$)	&	4.750144752208($-5$)	&	60	&	7.941482646732($-1$)	&	2.058517353268($-1$)	\\
2	&	9.998099635570($-1$)	&	1.900364429836($-4$)	&	61	&	7.859882307666($-1$)	&	2.140117692334($-1$)	\\
3	&	9.995723030365($-1$)	&	4.276969635440($-4$)	&	62	&	7.776115797026($-1$)	&	2.223884202974($-1$)	\\
4	&	9.992393636340($-1$)	&	7.606363659938($-4$)	&	63	&	7.690147901550($-1$)	&	2.309852098450($-1$)	\\
5	&	9.988109305211($-1$)	&	1.189069478858($-3$)	&	64	&	7.601943891394($-1$)	&	2.398056108606($-1$)	\\
6	&	9.982867272690($-1$)	&	1.713272730966($-3$)	&	65	&	7.511469665927($-1$)	&	2.488530334073($-1$)	\\
7	&	9.976664156839($-1$)	&	2.333584316116($-3$)	&	66	&	7.418691912491($-1$)	&	2.581308087509($-1$)	\\
\hline
\end{tabular}
\end{table}

\newpage

\begin{table}
\footnotesize
\caption{\small{Relative dia- and paramagnetic contributions to magnetic shielding constants  of hydrogenlike atoms in the excited state $3p_{1/2}$, obtained with $\alpha^{-1}=137.035 \: 999 \: 177$ (CODATA 2024).}}
\label{tab:3p_1-2}
\begin{tabular}{rccrcc}
\hline
$Z$ & $\sigma_{d}/ \sigma$ & $\sigma_{p}/ \sigma$ & Z  & $\sigma_{d}/ \sigma$ &  $\sigma_{p}/ \sigma$ \\
\hline
1	&	1.331285830381($-5$)	&	9.999866871417($-1$)	&	60	&	4.809700861035($-2$)	&	9.519029913896($-1$)	\\
2	&	5.325166948561($-5$)	&	9.999467483305($-1$)	&	61	&	4.971352659847($-2$)	&	9.502864734015($-1$)	\\
3	&	1.198171421248($-4$)	&	9.998801828579($-1$)	&	62	&	5.135603473602($-2$)	&	9.486439652640($-1$)	\\
4	&	2.130104564098($-4$)	&	9.997869895436($-1$)	&	63	&	5.302437889659($-2$)	&	9.469756211034($-1$)	\\
5	&	3.328332629572($-4$)	&	9.996671667370($-1$)	&	64	&	5.471838391116($-2$)	&	9.452816160888($-1$)	\\
6	&	4.792876811265($-4$)	&	9.995207123189($-1$)	&	65	&	5.643785100550($-2$)	&	9.435621489945($-1$)	\\
7	&	6.523762968007($-4$)	&	9.993476237032($-1$)	&	66	&	5.818255491366($-2$)	&	9.418174450863($-1$)	\\
\hline
\end{tabular}
\end{table}

\clearpage

\thispagestyle{empty}

\begin{table}
\footnotesize
\caption{\small{Relative dia- and paramagnetic contributions to magnetic shielding constants  of hydrogenlike atoms in the excited state $3p_{3/2}$ ($\mu=\pm 1/2$), obtained with $\alpha^{-1}=137.035 \: 999 \: 177$ (CODATA 2024).}}
\label{tab:3p_3-2_mu_1-2}
\begin{tabular}{rccrcc}
\hline
$Z$ & $\sigma_{d}/ \sigma$ & $\sigma_{p}/ \sigma$ & Z  & $\sigma_{d}/ \sigma$ &  $\sigma_{p}/ \sigma$ \\
\hline
1	&	-1.065044733703($-5$)	&	1.000010650447($0$)	&	70	&	1.056872965322($0$)	&	1.138700966968($0$)	\\
2	&	-4.260390688131($-5$)	&	1.000042603907($0$)	&	71	&	1.058663639312($0$)	&	1.143458746234($0$)	\\
3	&	-9.586673246555($-5$)	&	1.000095866732($0$)	&	72	&	1.060489537593($0$)	&	1.148337697796($0$)	\\
4	&	-1.704495179229($-4$)	&	1.000170449518($0$)	&	73	&	1.062351173324($0$)	&	1.153341133342($0$)	\\
5	&	-2.663671032599($-4$)	&	1.000266367103($0$)	&	74	&	1.064249075161($0$)	&	1.158472512288($0$)	\\
6	&	-3.836385833094($-4$)	&	1.000383638583($0$)	&	75	&	1.066183787770($0$)	&	1.163735449773($0$)	\\
7	&	-5.222873188729($-4$)	&	1.000522287319($0$)	&	76	&	1.068155872372($0$)	&	1.169133725193($0$)	\\
\hline
\end{tabular}
\end{table}

\clearpage

\thispagestyle{empty}

\begin{table}
\footnotesize
\caption{\small{Relative dia- and paramagnetic contributions to magnetic shielding constants  of hydrogenlike atoms in the excited state $3p_{3/2}$ ($\mu=\pm 3/2$), obtained with $\alpha^{-1}=137.035 \: 999 \: 177$ (CODATA 2024).}}
\label{tab:3p_3-2_mu_3-2}
\begin{tabular}{rccrcc}
\hline
$Z$ & $\sigma_{d}/ \sigma$ & $\sigma_{p}/ \sigma$ & Z  & $\sigma_{d}/ \sigma$ &  $\sigma_{p}/ \sigma$ \\
\hline
1	&	9.999810069992($-1$)	&	1.899300079631($-5$)	&	70	&	9.066431610843($-1$)	&	9.335683891566($-2$)	\\
2	&	9.999240277844($-1$)	&	7.597221564926($-5$)	&	71	&	9.039523133461($-1$)	&	9.604768665394($-2$)	\\
3	&	9.998290617185($-1$)	&	1.709382815493($-4$)	&	72	&	9.012232922825($-1$)	&	9.877670771754($-2$)	\\
4	&	9.996961077412($-1$)	&	3.038922588084($-4$)	&	73	&	8.984561192795($-1$)	&	1.015438807205($-1$)	\\
5	&	9.995251643709($-1$)	&	4.748356290512($-4$)	&	74	&	8.956508175737($-1$)	&	1.043491824263($-1$)	\\
6	&	9.993162297076($-1$)	&	6.837702924023($-4$)	&	75	&	8.928074123208($-1$)	&	1.071925876792($-1$)	\\
7	&	9.990693014361($-1$)	&	9.306985638685($-4$)	&	76	&	8.899259306644($-1$)	&	1.100740693356($-1$)	\\
\hline
\end{tabular}
\end{table}

\clearpage

\thispagestyle{empty}

\begin{table}
\footnotesize
\caption{\small{Relative dia- and paramagnetic contributions to magnetic shielding constants  of hydrogenlike atoms in the excited state $3d_{3/2}$ ($\mu=\pm 1/2$), obtained with $\alpha^{-1}=137.035 \: 999 \: 177$ (CODATA 2024).}}
\label{tab:3d_3-2_mu_1-2}
\begin{tabular}{rccrcc}
\hline
$Z$ & $\sigma_{d}/ \sigma$ & $\sigma_{p}/ \sigma$ & Z  & $\sigma_{d}/ \sigma$ &  $\sigma_{p}/ \sigma$ \\
\hline
1	&	1.643543034371($-5$)	&	9.999835645697($-1$)	&	70	&	7.655510564127($-2$)	&	9.234448943587($-1$)	\\
2	&	6.573964687731($-5$)	&	9.999342603531($-1$)	&	71	&	7.864507447376($-2$)	&	9.213549255262($-1$)	\\
3	&	1.479064266699($-4$)	&	9.998520935733($-1$)	&	72	&	8.075849903838($-2$)	&	9.192415009616($-1$)	\\
4	&	2.629254000415($-4$)	&	9.997370746000($-1$)	&	73	&	8.289513845562($-2$)	&	9.171048615444($-1$)	\\
5	&	4.107820533691($-4$)	&	9.995892179466($-1$)	&	74	&	8.505474971838($-2$)	&	9.149452502816($-1$)	\\
6	&	5.914577330119($-4$)	&	9.994085422670($-1$)	&	75	&	8.723708772464($-2$)	&	9.127629122754($-1$)	\\
7	&	8.049296503528($-4$)	&	9.991950703496($-1$)	&	76	&	8.944190530929($-2$)	&	9.105580946907($-1$)	\\
\hline
\end{tabular}
\end{table}

\clearpage

\thispagestyle{empty}

\begin{table}
\footnotesize
\caption{\small{Relative dia- and paramagnetic contributions to magnetic shielding constants  of hydrogenlike atoms in the excited state $3d_{3/2}$ ($\mu=\pm 3/2$), obtained with $\alpha^{-1}=137.035 \: 999 \: 177$ (CODATA 2024).}}
\label{tab:3d_3-2_mu_3-2}
\begin{tabular}{rccrcc}
\hline
$Z$ & $\sigma_{d}/ \sigma$ & $\sigma_{p}/ \sigma$ & Z  & $\sigma_{d}/ \sigma$ &  $\sigma_{p}/ \sigma$ \\
\hline
1	&	3.697895855246($-5$)	&	9.999630210414($-1$)	&	70	&	1.558081624632($-1$)	&	8.441918375368($-1$)	\\
2	&	1.479020507597($-4$)	&	9.998520979492($-1$)	&	71	&	1.595997878873($-1$)	&	8.404002121127($-1$)	\\
3	&	3.327279336823($-4$)	&	9.996672720663($-1$)	&	72	&	1.634088260678($-1$)	&	8.365911739322($-1$)	\\
4	&	5.913877271157($-4$)	&	9.994086122729($-1$)	&	73	&	1.672339809709($-1$)	&	8.327660190291($-1$)	\\
5	&	9.237850506000($-4$)	&	9.990762149494($-1$)	&	74	&	1.710739524373($-1$)	&	8.289260475627($-1$)	\\
6	&	1.329796075169($-3$)	&	9.986702039248($-1$)	&	75	&	1.749274365337($-1$)	&	8.250725634663($-1$)	\\
7	&	1.809269589777($-3$)	&	9.981907304102($-1$)	&	76	&	1.787931258985($-1$)	&	8.212068741015($-1$)	\\
\hline
\end{tabular}
\end{table}

\clearpage

\thispagestyle{empty}

\begin{table}
\footnotesize
\caption{\small{Relative dia- and paramagnetic contributions to magnetic shielding constants  of hydrogenlike atoms in the excited state $3d_{5/2}$ ($\mu=\pm 1/2$), obtained with $\alpha^{-1}=137.035 \: 999 \: 177$ (CODATA 2024).}}
\label{tab:3d_5-2_mu_1-2}
\begin{tabular}{rccrcc}
\hline
$Z$ & $\sigma_{d}/ \sigma$ & $\sigma_{p}/ \sigma$ & Z  & $\sigma_{d}/ \sigma$ &  $\sigma_{p}/ \sigma$ \\
\hline
1	&	-1.584891546651($-5$)	&	1.000015848915($0$)	&	70	&	-8.557998213307($-2$)	&	1.085579982133($0$)	\\
2	&	-6.339923900256($-5$)	&	1.000063399239($0$)	&	71	&	-8.830300780915($-2$)	&	1.088303007809($0$)	\\
3	&	-1.426617041313($-4$)	&	1.000142661704($0$)	&	72	&	-9.108123501932($-2$)	&	1.091081235019($0$)	\\
4	&	-2.536542071056($-4$)	&	1.000253654207($0$)	&	73	&	-9.391552581371($-2$)	&	1.093915525814($0$)	\\
5	&	-3.964018174880($-4$)	&	1.000396401817($0$)	&	74	&	-9.680676777147($-2$)	&	1.096806767771($0$)	\\
6	&	-5.709367929791($-4$)	&	1.000570936793($0$)	&	75	&	-9.975587481056($-2$)	&	1.099755874811($0$)	\\
7	&	-7.772985985104($-4$)	&	1.000777298599($0$)	&	76	&	-1.027637880313($-1$)	&	1.102763788031($0$)	\\
\hline
\end{tabular}
\end{table}

\clearpage

\thispagestyle{empty}

\begin{table}
\footnotesize
\caption{\small{Relative dia- and paramagnetic contributions to magnetic shielding constants  of hydrogenlike atoms in the excited state $3d_{5/2}$ ($\mu=\pm 3/2$), obtained with $\alpha^{-1}=137.035 \: 999 \: 177$ (CODATA 2024).}}
\label{tab:3d_5-2_mu_3-2}
\begin{tabular}{rccrcc}
\hline
$Z$ & $\sigma_{d}/ \sigma$ & $\sigma_{p}/ \sigma$ & Z  & $\sigma_{d}/ \sigma$ &  $\sigma_{p}/ \sigma$ \\
\hline
1	&	-2.905672879132($-5$)	&	1.000029056729($0$)	&	70	&	-1.697502613333($-1$)	&	1.169750261333($0$)	\\
2	&	-1.162380796426($-4$)	&	1.000116238080($0$)	&	71	&	-1.756349753076($-1$)	&	1.175634975308($0$)	\\
3	&	-2.615775585728($-4$)	&	1.000261577559($0$)	&	72	&	-1.816741484412($-1$)	&	1.181674148441($0$)	\\
4	&	-4.651310509303($-4$)	&	1.000465131051($0$)	&	73	&	-1.878722547597($-1$)	&	1.187872254760($0$)	\\
5	&	-7.269768844612($-4$)	&	1.000726976884($0$)	&	74	&	-1.942339760268($-1$)	&	1.194233976027($0$)	\\
6	&	-1.047215917791($-3$)	&	1.001047215918($0$)	&	75	&	-2.007642134210($-1$)	&	1.200764213421($0$)	\\
7	&	-1.425971654510($-3$)	&	1.001425971655($0$)	&	76	&	-2.074681000237($-1$)	&	1.207468100024($0$)	\\
\hline
\end{tabular}
\end{table}

\clearpage

\thispagestyle{empty}

\begin{table}
\footnotesize
\caption{\small{Relative dia- and paramagnetic contributions to magnetic shielding constants  of hydrogenlike atoms in the excited state $3d_{5/2}$ ($\mu=\pm 5/2$), obtained with $\alpha^{-1}=137.035 \: 999 \: 177$ (CODATA 2024).}}
\label{tab:3d_5-2_mu_5-2}
\begin{tabular}{rccrcc}
\hline
$Z$ & $\sigma_{d}/ \sigma$ & $\sigma_{p}/ \sigma$ & Z  & $\sigma_{d}/ \sigma$ &  $\sigma_{p}/ \sigma$ \\
\hline
1	&	9.999877981436($-1$)	&	1.220185642468($-5$)	&	70	&	9.418368744155($-1$)	&	5.816312558447($-2$)	\\
2	&	9.999511934092($-1$)	&	4.880659078654($-5$)	&	71	&	9.402100348200($-1$)	&	5.978996518002($-2$)	\\
3	&	9.998901883015($-1$)	&	1.098116984883($-4$)	&	72	&	9.385627206274($-1$)	&	6.143727937260($-2$)	\\
4	&	9.998047869943($-1$)	&	1.952130056647($-4$)	&	73	&	9.368950355641($-1$)	&	6.310496443587($-2$)	\\
5	&	9.996949953301($-1$)	&	3.050046698785($-4$)	&	74	&	9.352070843953($-1$)	&	6.479291560470($-2$)	\\
6	&	9.995608208189($-1$)	&	4.391791810932($-4$)	&	75	&	9.334989729095($-1$)	&	6.650102709047($-2$)	\\
7	&	9.994022726371($-1$)	&	5.977273629247($-4$)	&	76	&	9.317708079036($-1$)	&	6.822919209640($-2$)	\\
\hline
\end{tabular}
\end{table}

\begin{table}
\footnotesize
\caption{\small{Comparison of present values (the upper entries) of the diamagnetic part $\sigma_d$ of magnetic shielding constants  (in units of $10^{-6}$) for selected Dirac one-electron atoms in the ground state (1s$_{1/2}$) and the first excited state, i.e: 2p$_{1/2}$ and 2p$_{3/2}$, with those obtained by Pyper and Zhang \cite{Pype99a} (the lower entries). All the numbers were obtained using $\alpha^{-1}=137.035 \: 989 \: 5$ (from CODATA 1986). The present results were computed from the analytical formula given by Eqs.\ (\ref{eq:2})--(\ref{eq:4}).}}
\label{tab:comparison2}
\begin{tabular}{rllll}
\hline
$Z$ & $1s_{1/2}$ & $2p_{1/2}$ & $2p_{3/2}$ ($\mu=\pm 1/2$)  & $2p_{3/2}$ ($\mu=\pm 3/2$) \\
\hline \\
1   & 17.75 045 398 967  & 4.437 672 576 230    & 3.550 090 797 933  & 5.325 136 196 900    \\
    & 17.8               & 4.4            &3.6          & 5.3               \\*[2ex]
11  & 195.2 549 938 863  & 48.89 263 463 239    & 39.05 099 877 727  & 58.57 649 816 590    \\
    & 195.3              &48.9             & 39.1           & 58.6                 \\*[2ex]
37  &656.7 667 976 177  & 167.2 984 531 435    & 131.3 533 595 235  & 197.0 300 392 853   \\
    & 656.8              & 167.3              & 131.4           & 197.0               \\*[2ex] 
55  & 976.2 749 694 317   & 254.7 793 038 102    & 195.2 549 938 863  & 292.8 824 908 295    \\
    & 976.3             & 254.8              & 195.3            & 292.9                \\*[2ex]
74  & 1313.533 595 235   & 356.6 162 880 871   & 262.7 067 190 471  & 394.0 600 785 706      \\
    & 1313.5             & 356.6              & 262.7           & 394.1                \\*[2ex]
92  & 1633.041 767 049   & 468.9 590 861 089    &  326.6 083 534 099  & 489.9 125 301 148   \\
    & 1633.0             & 468.9              & 326.6           &489.9                \\
\\
\hline
\end{tabular}
\end{table}

\begin{table}
\footnotesize
\caption{\small{Comparison of present values (the upper entries) of the paramagnetic part $\sigma_p$ of magnetic shielding constants  (in units of $10^{-6}$) for selected Dirac one-electron atoms in the ground state (1s$_{1/2}$) and the first excited state, i.e: 2p$_{1/2}$ and 2p$_{3/2}$, with those obtained by Pyper and Zhang \cite{Pype99a} (the lower entries). All the numbers were obtained using $\alpha^{-1}=137.035 \: 989 \: 5$ (from CODATA 1986). The present results were computed from the analytical formula given by Eqs.\ (\ref{eq:2})--(\ref{eq:4}).}}
\label{tab:comparison3}
\begin{tabular}{rllll}
\hline
$Z$ & $1s_{1/2}$ & $2p_{1/2}$ & $2p_{3/2}$ ($\mu=\pm 1/2$)  & $2p_{3/2}$ ($\mu=\pm 3/2$) \\
\hline \\
1   & 254.7 087 540 815 $\times 10^{-5}$  & 148147.1 621 021    & $-$148147.1 619 923  & 15.19 490 166 962 $\times 10^{-5}$   \\
    & 256 $\times 10^{-5}$                        & 148147.1 999 610            &$-$148147.2 00 0080          & 15 $\times 10^{-5}$               \\*[2ex]
11  & 3.422 742 348 979  & 13457.28 781 693    & $-$13457.13 995 737  & 0.2026 207 637 545    \\
    & 3.4             & 13457.25             &  $-$13457.11                            & 0.2                 \\*[2ex]
37  &144.6 232 009 308  & 3972.916 506 558    & $-$3966.509 047 635  & 7.863 199 259 112  \\
    & 144.6              & 3972.9             & $-$3966.5           & 7.8             \\*[2ex] 
55  & 555.1 816 954 109   & 2661.492 865 308    & $-$2635.972 535 267 & 26.52 113 424 921    \\
    & 555.2             &  2661.5              & $-$2636            & 26.5               \\*[2ex]
74  & 1788.531 850 896   & 2007.999 788 708  & $-$1920.449 568 192  & 67.26 349 470 640      \\
    & 1788.4             & 2008              & $-$1920.5           & 67.2                \\*[2ex]
92  & 5518.411 839 600   & 1799.139 473 066    &  $-$1502.291 749 952  & 136.0 898 745 309   \\
    & 5518.0             &  1799.1              & $-$1502.3           &136.1               \\
\\
\hline
\end{tabular}
\end{table}

\newpage

\begin{landscape}

\begin{table}
\footnotesize
\caption{\small{Relativistic total magnetic shielding constants $\sigma=\sigma_{d}+\sigma_{p}$ (in the units of $\alpha^2$) for states $1s_{1/2}$, $2s_{1/2}$, $2p_{1/2}$ and $2p_{3/2}$ of selected  hydrogenlike ions, obtained with $\alpha^{-1}=137.035 \: 999 \: 139$ (CODATA 2014) [the upper entries],  with $\alpha^{-1}=137.035 \: 999 \: 084$ (CODATA 2018) [the middle entries], and with $\alpha^{-1}=137.035 \: 999 \: 177$ (CODATA 2024) [the lower entries].}}
\label{tab:comparison}
\begin{tabular}{rlllll}
\hline
$Z$ & $1s_{1/2}$ & $2s_{1/2}$ & $2p_{1/2}$ & $2p_{3/2}$ ($\mu=\pm 1/2$)  & $2p_{3/2}$ ($\mu=\pm 3/2$) \\
\hline \\
   1  & 3.333 811 647 331($-$1)   & 8.334 039 107 064($-$2)  & 2.782 118 900 586(3) & $-$2.781 968 897 415(3)  & 1.000 028 534 293($-$1)  \\
     & 3.333 811 647 331($-$1) & 8.334 039 107 064($-$2) & 2.782 118 898 353(3)  &  $-$2.781 968 895 182(3)  & 1.000 028 534 293($-$1) \\
		&  3.333 811 647 331($-1$) & 8.334 039 107 063($-2$) & 2.782 118 902 129(3) &   $-$2.781 968 898 958(3)  & 1.000 028 534 293($-1$) \\*[2ex] 
   2  & 6.670 494 089 568($-$1)  & 1.667 231 453 748($-$1)  & 1.391 156 693 555(3) & $-$1.390 856 668 182(3)  & 2.000 228 284 951($-$1) \\
     & 6.670 494 089 571($-$1)  & 1.667 231 453 749($-$1) & 1.391 156 692 438(3)  & $-$1.390 856 667 065(3) & 2.000 228 284 951($-$1) \\
		&  6.670 494 089 566($-1$) &  1.667 231 453 748($-1$) & 1.391 156 694 326(3)  & $-$1.390 856 668 953(3) & 2.000 228 284 951($-1$) \\*[2ex] 
 5    & 1.672 656 996 173(0)  & 4.175 509 888 277($-$1) & 5.567 352 419 207(2) & $-$5.559 848 448 573(2)  & 5.003 568 113 104($-$1) \\
     & 1.672 656 996 177(0) & 4.175 509 888 284($-$1) &  5.567 352 414 741(2) & $-$5.559 848 444 107(2)  & 5.003 568 113 107($-$1) \\
		&  1.672 656 996 169(0) & 4.175 509 888 272($-1$) &  5.567 352 422 293(2) & $-$5.559 848 451 659(2) &  5.003 568 113 102($-1$) \\*[2ex] 
    10 & 3.381 543 223 662(0) &  8.404 609 846 285($-$1) & 2.788 563 837 135(2)  &  $-$2.773 531 895 705(2)  & 1.002 857 811 697(0) \\
     & 3.381 543 223 701(0)  & 8.404 609 846 343($-$1) & 2.788 563 834 901(2)  & $-$2.773 531 893 472(2)  & 1.002 857 811 699(0) \\
		&  3.381 543 223 635(0) &  8.404 609 846 245($-1$) & 2.788 563 838 677(2) &  $-$2.773 531 897 248(2) &  1.002 857 811 695(0) \\*[2ex] 
   20  &  7.061 805 272 665(0) & 1.725 442 452 936(0)  & 1.404 221 872 018(2)  & $-$1.373 960 527 835(2) & 2.022 969 375 495(0) \\
     &  7.061 805 272 993(0) & 1.725 442 452 985(0) & 1.404 221 870 902(2) & $-$1.373 960 526 719(2)  &  2.022 969 375 513(0) \\
		&   7.061 805 272 439(0) & 1.725 442 452 902(0) & 1.404 221 872 789(2) & $-$1.373 960 528 607(2) &   2.022 969 375 482(0) \\*[2ex] 
   40  & 1.683 584 880 967(1)  & 3.867 902 201 437(0) & 7.235 008 656 373(1)   & $-$6.611 994 893 890(1)  & 4.187 254 265 134(0) \\
     & 1.683 584 881 289(1)   & 3.867 902 201 943(0) & 7.235 008 650 809(1)  & $-$6.611 994 888 305(1) & 4.187 254 265 288(0) \\
		&  1.683 584 880 745(1) &   3.867 902 201 087(0) & 7.235 008 660 217(1) &  $-$6.611 994 897 748(1) & 4.187 254 265 027(0) \\*[2ex] 
  50   & 2.410 407 187 488(1) & 5.325 805 020 255(0) & 5.932 084 204 416(1)   & $-$5.133 432 930 687(1) & 5.371 024 936 168(0) \\
     & 2.410 407 188 229(1) & 5.325 805 021 466(0)  & 5.932 084 199 994(1)  &  $-$5.133 432 926 217(1)  & 5.371 024 936 478(0) \\
		 & 2.410 407 186 976(1) & 5.325 805 019 419(0) &  5.932 084 207 472(1) &   $-$5.133 432 933 775(1) &  5.371 024 935 954(0) \\*[2ex] 
    80 &  7.482 499 069 208(1) & 1.505 516 507 875(1) & 4.286 785 449 937(1)   & $-$2.775 434 972 921(1) &  9.621 170 168 023(0) \\
     & 7.482 499 076 656(1)  & 1.505 516 509 318(1) &  4.286 785 447 607(1) & $-$2.775 434 970 111(1) & 9.621 170 169 467(0)\\
		 & 7.482 499 064 061(1) &  1.505 516 506 878(1) &  4.286 785 451 546(1) & $-$2.775 434 974 862(1) & 9.621 170 167 026(0) \\*[2ex] 
   100  &  2.224 303 767 330(2)  & 4.645 586 300 011(1) & 4.622 613 420 492(1) & $-$1.887 430 743 694(1)  & 1.337 342 161 028(1)  \\
     &  2.224 303 772 917(2) &  4.645 586 312 774(1) & 4.622 613 421 944(1)   & $-$1.887 430 741 421(1) & 1.337 342 161 348(1) \\
		&   2.224 303 763 470(2) &  4.645 586 291 193(1) & 4.622 613 419 489(1) &   $-$1.887 430 745 264(1) & 1.337 342 160 807(1) \\*[2ex] 
   118  & 9.084 394 854 718(3)   & 2.355 074 407 820(3)  & 6.641 887 213 111(2)   & $-$1.277 245 285 008(1) & 1.776 612 163 038(1)  \\
     & 9.084 395 498 304(3)  &  2.355 074 576 922(3)   & 6.641 887 666 149(2)   & $-$1.277 245 283 040(1) & 1.776 612 163 646(1) \\
		&  9.084 394 410 059(3) &   2.355 074 290 987(3) &   6.641 886 900 104(2) &   $-$1.277 245 286 368(1) & 1.776 612 162 617(1) \\*[2ex]    
    130 &   &   &   &  $-$9.263 105 717 640(0) & 2.146 615 815 223(1) \\
     &  & &  & $-$9.263 105 699 332(0)  & 2.146 615 816 138(1) \\
		& & & &    $-$9.263 105 730 288(0) &  2.146 615 814 590(1) \\*[2ex] 
   137  &  & &  &  $-$7.346 108 509 931(0)   & 2.400 367 487 661(1)  \\
     &  & &  & $-$7.346 108 492 209(0)  & 2.400 367 488 819(1)   \\
		& & & &    $-$7.346 108 522 175(0)  & 2.400 367 486 861(1) \\
\\
\hline
\end{tabular}
\end{table}

\end{landscape}

\newpage

%
%

%
\end{document}